\documentclass[11pt]{amsart}

\usepackage{amscd}
\usepackage{amsmath}
\usepackage{graphicx}
\usepackage{amsfonts}
\usepackage{amssymb}
\begin{document}

\title[Positive linear maps and criterion of entanglement]
{A characterization of positive linear maps and  criteria of
entanglement for quantum states}

\author{Jinchuan Hou}
\address{Department of
Mathematics\\
Taiyuan University of Technology\\
 Taiyuan 030024,
  P. R. of China}
\email{jinchuanhou@yahoo.com.cn, houjinchuan@tyut.edu.cn}


\thanks{This work is partially supported by National Natural Science
Foundation of China (No. 10771157) and Research Grant to Returned
Scholars of Shanxi (2007-38).}

\begin{abstract}

Let  $H$ and $K$ be   (finite or infinite dimensional) complex
Hilbert spaces. A characterization of positive completely bounded
normal linear maps from ${\mathcal B}(H)$ into ${\mathcal B}(K)$ is
given, which particularly gives a characterization of positive
elementary operators including all positive linear maps between
matrix algebras. This characterization is then applied give a
representation of quantum channels (operations) between
infinite-dimensional systems. A necessary and sufficient criterion
of separability is give which shows that a state $\rho$ on $H\otimes
K$ is separable if and only if $(\Phi\otimes I)\rho\geq 0$ for all
positive finite rank elementary operators $\Phi$. Examples of NCP
and indecomposable positive linear maps are given and are used to
recognize  some entangled states that cannot be recognized by the
PPT criterion and the realignment criterion.

\end{abstract}
\maketitle

\section{Introduction}

Positive linear maps and completely positive linear maps are found
to be very important in quantum mechanics, quantum computation and
quantum information. In fact they can be used to recognize entangled
states, and every quantum channel is represented as a trace
preserving completely positive linear map.

In quantum mechanics, a quantum system is
 associated with a separable complex Hilbert space $H$, i.e.,
the state space.  A quantum state is described as a density operator
$\rho\in{\mathcal T}(H)\subseteq{\mathcal B}(H)$ which is positive
and has trace 1, where ${\mathcal B}(H)$ and ${\mathcal T}(H)$
denote the von Neumann algebras of all bounded linear operators and
the trace-class of all operators $T$ with $\|T\|_1={\rm
Tr}((T^\dagger T)^{\frac{1}{2}})<\infty$, respectively.  $\rho$ is a
pure state if $\rho^2=\rho$; $\rho$ is a mixed state if
$\rho^2\not=\rho$. The state space $H$ of a composite quantum system
is the tensor product of the state spaces of the component quantum
systems $H_i$, that is $H=H_1\otimes H_2\otimes \ldots \otimes H_k$.
In this paper we are mainly interested in bipartite systems, that
is, the case $k=2$. Let $H$ and $K$ be finite dimensional and let
$\rho$ be a state acting on $H\otimes K$. $\rho$ is said to be
separable if $\rho$ can be written as
$$\rho=\sum_{i=1}^k p_i \rho_i\otimes \sigma _i,$$
where $\rho_i$ and $\sigma_i$ are states on $H$ and $K$
respectively, and $p_i$ are positive numbers with $\sum
_{i=1}^kp_i=1$. Otherwise, $\rho$ is said to be inseparable or
entangled (ref. \cite{BZ, NC}). For the case that at least one of
$H$ and $K$ is of infinite dimension,  by  Werner \cite{W},  a state
$\rho$ acting on $H\otimes K$ is called separable if it can be
approximated in the trace norm by the states of the form
$$\sigma=\sum_{i=1}^n p_i \rho_i\otimes \sigma _i,$$
where $\rho_i$ and $\sigma_i$ are states on $H$ and $K$
respectively, and $p_i$ are positive numbers with
$\sum_{i=1}^np_i=1$. Otherwise, $\rho$ is called an entangled state.

The quantum entangled states have been used as basic resources in
quantum information processing and communication (see
\cite{NC,BB,BP,DE,DE1,S}). Generally, to decide whether or not a
state of composite quantum systems is entangled is one of the most
challenging task of this field \cite{NC}. For the case of $2\times
2$ or $2\times 3$ systems, that is, for the case $\dim H=\dim K=2$
or $\dim H=2, \ \dim K=3$, a state is separable if and only if it is
a PPT (Positive Partial Transpose) state \cite{Hor, Pe}. But PPT is
only a necessary condition for a state to be separable acting on
Hilbert space of higher dimensions. There are PPT states that are
entangled. It is known that PPT entangled states belong to the class
of bound entangled states \cite{HHH}. In \cite{CW}, the realignment
criterion for separability in finite dimensional systems was found.
It is independent of the PPT criterion and  can detect some bound
entangled states that cannot be recognized by the PPT criterion.
There are several  other sufficient criteria for entanglement such
as the reduction criteria \cite{Hor2, CAG,HHH1}.

A most general approach to study the entanglement of quantum states
in finite dimensional systems is based on the notion of entanglement
witnesses (see \cite{Hor}). A Hermitian operator $W$ acting on
$H\otimes K$ is said to be an entanglement witness (briefly, EW), if
$W$ is not positive and ${\rm Tr}(W\sigma)\geq 0$ holds for all
separable states $\sigma$. Thus, $\rho$ is entangled if and only if
there exists an EW $W$ such that ${\rm Tr}(W\rho) < 0$ \cite{Hor}.
This entanglement witness criterion is also valid for infinite
dimensional systems.  Clearly, constructing entanglement witnesses
is a hard task. A recent result in \cite{HQ} states that every
entangled state in a bipartite (finite or infinite dimensional)
system can be detected by a witness of the form $cI-F$, where $c$ is
a nonnegative number and $F$ is a finite rank self-adjoint operator.

Another general approach to detect entanglement is based on positive
maps. It is obvious that if $\rho$ is a state on $H\otimes K$, then
for every completely positive (briefly, CP) linear map $\Phi
:{\mathcal B}(H)\rightarrow {\mathcal B}(K)$, the operator
$(\Phi\otimes I)\rho\in{\mathcal B}(K\otimes K)$ is always positive;
if $\rho$ is separable, then for every
 positive linear map $\Phi :{\mathcal B}(H)\rightarrow
{\mathcal B}(K)$, the operator $(\Phi\otimes I)\rho$ is always
positive on $K\otimes K$ (or, for every
 positive linear map $\Phi :{\mathcal B}(K)\rightarrow
{\mathcal B}(H)$, the operator $(I\otimes \Phi)\rho$ is always
positive on $H\otimes H$). The converse of the last statement is
also true. In \cite{Hor}, it was shown that

{\bf Horodeckis' Theorem}. \cite[Theorem 2]{Hor} {\it Let $H$, $K$
be finite dimensional complex Hilbert spaces and $\rho$ be a state
acting on $H\otimes K$. Then $\rho$ is separable if and only if for
any positive linear map $\Phi :{\mathcal B}(H)\rightarrow {\mathcal
B}(K)$, the operator $(\Phi\otimes I)\rho $ is positive on $K\otimes
K$.}

The positive map criterion and the witness criterion for
entanglement are two of few known necessary and sufficient criteria.
These two criteria are closely connected by the so-called
 the Jamio{\l}kowski-Choi isomorphism \cite{Hor, SS, SSZ, GKM}.
 Recall that a positive map is said to be decomposable if and only
 if it is the sum of a CP map and a map which is the transpose of
 some CP map. It is obvious that a decomposable positive map can not
 detect any PPT entangled states \cite{K}.

Let us consider the case that at least one of $H$ and $K$ is of
infinite dimension. As every positive linear map $\eta$ between von
Neumann algebras is bounded and $\|\eta\|=\|\eta(I)\|$ (see
\cite[Exercise 10.5.10]{Kad}),  $\rho$ is separable on $H\otimes K$
still implies that, for any completely bounded positive linear map
$\Phi:{\mathcal B}(H)\rightarrow{\mathcal B}(K)$, the operator
$(\Phi\otimes I)\rho$ is positive on $K\otimes K$. The
infinite-dimensional version of Horodeckis' Theorem above was
obtained by St${\o}$rmer \cite{Sto}. Recall that a positive linear
map $\Phi :{\mathcal B}(H)\rightarrow{\mathcal B}(K)$  is said to be
normal if it is weakly continuous on bounded sets, or equivalently,
if it is ultra-weakly continuous (i.e., if $\{A_\alpha\}$ is a
bounded net and there is $A\in{\mathcal B}(H)$ such that $\langle
x|A_\alpha|y\rangle$ converges to $\langle y|A|x\rangle$ for any
$|x\rangle\in H, |y\rangle\in K$, then $\langle
x|\Phi(A_\alpha)|y\rangle$ converges to $\langle y|\Phi(A)|x\rangle$
for any $|x\rangle\in H, |y\rangle\in K$. ref. \cite[pp.59]{Dix}).

{\bf St$\o$rmer's Theorem}. \cite{Sto} {\it Let $H,K$ be Hilbert
spaces, $\rho$ be a state acting on $H\otimes K$. Then $\rho$ is
separable if and only if for any normal positive linear map $\Phi
:{\mathcal B}(H)\rightarrow {\mathcal B}(K)$, the operator
$(\Phi\otimes I)\rho $ is positive on $K\otimes K$.}

Thus,   for a state $\rho$ on $H\otimes K$, if there exists a normal
positive map $\Phi :{\mathcal B}(H)\rightarrow {\mathcal B}(K)$ such
that $(\Phi\otimes I)\rho$ is not positive or unbounded, then $\rho$
is entangled. In this situation, $\Phi$ can never be completely
positive. Therefore, to detect the inseparability of states, the key
is to find the normal positive linear maps that are not completely
positive. In the case that $\dim H=\dim K =n$, the transpose
$A\mapsto A^T$ and the map $A\mapsto {\rm Tr}(A)I-A$ are well known
positive maps that are not completely positive.

Positive linear maps  have attracted much attention of physicists
working in quantum information science in recent decades because of
Horodeckis' positive map criterion. Great efforts have be payed to
find as many as possible positive maps that are not CP, and then use
them to detect some entangled states \cite{HHH1,CK,CK2, AS}, for
finite dimensional systems. Positive linear maps and completely
positive linear maps are also important mathematical topics studied
intensively in a general setting of C$^*$-algebras by
mathematicians. The completely positive linear maps can be
understood quite well. However, the structure of positive linear
maps  is drastically nontrivial even for the finite dimensional case
(\cite{C1}-\cite{P}).

Note that every linear map $\Phi$ from $\mathcal B(H)$ into
$\mathcal B(K)$ is an elementary operator if both $H$ and $K$ are
finite dimensional, that is, there exist operators $A_1, A_2, \ldots
, A_k\in{\mathcal B}(H,K)$ and $B_1, B_2, \ldots , B_k\in{\mathcal
B}(K,H)$, such that $\Phi (T)=\sum_{i=1}^kA_iTB_i$ for all
$T\in{\mathcal B}(H)$. So, it is also  basic important and
interesting to find as many as possible characterizations of
positive elementary operators and characterizations of completely
positive elementary operators, and then, to apply them to get some
criteria for the entanglement of states.

A characterization of positive elementary operators was obtained in
\cite{H4} in terms of contractively locally linear combinations.
This is the only known necessary and sufficient condition for an
elementary operator to be positive. The purpose of this paper is to
give a characterization of positive completely bounded normal maps
between $\mathcal B(H)$ and $\mathcal B(K)$, which including all
positive elementary operators. Consequently, we obtain concrete
representations of the completely bounded linear maps, positive
completely bounded linear maps and completely positive linear maps
between the trace-classes ${\mathcal T}(H)$ and ${\mathcal T}(K)$,
which allow us to obtain a representation of quantum operations
channels (operations) for infinite-dimensional systems. Apply our
characterization of positive maps that are not CP, a necessary and a
sufficient criterion, that is, the elementary operator criterion of
separability is proved. Finally, some positive elementary operators
are constructed so that they are not completely positive, even
indecomposable, and then used to recognize some entangled quantum
states that cannot be detected by the PPT criterion and the
realignment criterion.

The paper is organized as follows. Section 2 is of mathematics. We
show that the set of  completely bounded normal linear maps
coincides with the set of generalized elementary operators in the
setting of separable Hilbert spaces,  and give a characterization of
positive (completely positive) generalized elementary operators
(Lemma 2.1, Theorem 2.5), which improve the main results in
\cite{H4}. Several simple necessary or sufficient conditions to sure
that a positive map is not completely positive are also provided. We
also show that the non-complete positivity of a positive elementary
operator is essentially determined by its behavior on
finite-dimensional subspaces. In Section 3, applying the results in
Section 2, some necessary and sufficient conditions for a completely
bounded linear map on the trace-class ${\mathcal T}(H)$ to be
positive, or to be completely positive are  given (see Theorem 3.2).
As a corollary, we get a  representation of quantum channels
(operations) for infinite dimensional systems (Corollary 3.3), which
is similar to that for finite dimensional systems. The  purpose of
Section 4 is to apply the results in Section 2 to get some criteria
of entangled quantum states both for finite-dimensional case and
infinite-dimensional case and deduce the main result of this paper,
i.e., the elementary operator criterion, valid for both finite
dimensional systems and infinite dimensional cases. We show that the
following statements are equivalent: (1) $\rho$ is separable; (2)
$(\Phi\otimes I)\rho\geq 0$ for every positive elementary operator
$\Phi$; (3) $(\Phi\otimes I)\rho\geq 0$ for every finite rank
positive elementary operator $\Phi$ (Theorem 4.5). Thus, a state
$\rho$ is entangled if and only if there exists an elementary
operator of the form $\Phi(\cdot )= \sum _{i=1}^{k}C_{i}(\cdot
)C_{i}^{\dagger}-\sum _{j=1}^{l}D_{j}(\cdot
)D_{j}^{\dagger}:{\mathcal B}(H)\rightarrow {\mathcal B}(K)$, where
 all $C_i$s and $D_j$s are of finite rank and $\{ D_1,\ldots , D_l\}$ is a
contractive locally linear combination of $\{C_1, \ldots ,C_k\}$,
such that the operator $(\Phi\otimes I)\rho $ is not positive. This
criterion improves the St$\o$mer' theorem greatly and is more
practical. Section 5 is devoted to illustrating how to apply the
results in Sections 2 and 4 to construct positive elementary
operators that are not completely positive and even indecomposable
(see propositions 5.1-5.2). These maps then are used to distinguish
some entangled states that cannot be recognized by PPT criterion as
well as the realignment criterion. In Section 6, we give a short
conclusion.

Throughout this paper, $H$ and $K$ are separable complex Hilbert
spaces that may be of infinite dimension  if no specific assumption
is made, and $\langle \cdot|\cdot\rangle$ stands for the inner
product in both of them. ${\mathcal B}(H,K)$ (${\mathcal B}(H)$ when
$K=H$) is the Banach space of all (bounded linear) operators from
$H$ into $K$. $A\in{\mathcal B}(H)$ is self-adjoint if $A=A^\dagger$
($A^\dagger$ stands for the adjoint operator of $A$); and $A$ is
positive, denoted by $A\geq 0$, if $A$ is self-adjoint with spectrum
falling in the interval $[0,\infty)$ (or equivalently, $\langle \psi
| A\psi\rangle\geq 0$ for all $|\psi\rangle\in H$). For any positive
integer $n$, $H^{(n)}$ denotes the direct sum of $n$ copies of $H$.
It is clear that every operator ${\bf A}\in{\mathcal
B}(H^{(n)},K^{(m)})$ can be written in an $n\times m$ operator
matrix ${\bf A}=(A_{ij})_{i,j}$ with $A_{ij}\in{\mathcal B}(H,K)$,
$i=1,2,\ldots,m$; $j=1,2,\ldots,n$. Equivalently, ${\mathcal
B}(H^{(n)},K^{(m)})$ is often written as ${\mathcal B}(H,K)\otimes
{\mathcal M}_{m\times n}({\mathbb C})$. We will write ${\bf A}^{\rm
T}=(A_{ij})^{\rm T}$ for the formal transpose matrix
$(A_{ji})_{i,j}$ of ${\bf A}$, ${\bf A}^{\rm t}=(A_{ji}^{\rm
t})_{i,j}$ for the usual transpose of $\bf A$, and denote by
$A^{(n)}$ the operator matrix $(A_{ij})\in{\mathcal
B}(H^{(n)},K^{(n)})$ with $A_{ii}=A$ and $A_{ij}=0$ if $i\not=j$. If
$\Phi$ is a linear map from ${\mathcal B}(H)$ into ${\mathcal
B}(K)$, we can define a linear map $\Phi_n:{\mathcal
B}(H^{(n)})\rightarrow{\mathcal B}(K^{(n)})$ by
$\Phi_n((A_{ij}))=(\Phi(A_{ij}))$. Recall that $\Phi$ is said to be
positive (resp. hermitian-preserving) if $A\in{\mathcal B}(H)$ is
positive (resp. self-adjoint) implies that $\Phi(A)$ is positive
(resp. self-adjoint). If $\Phi_n$ is positive we say $\Phi$ is
$n$-positive; if $\Phi_n$ is positive for every integer $n>0$, we
say that $\Phi$ is completely positive. Obviously, $\Phi$ is
completely positive $\Rightarrow$ $\Phi$ is positive $\Rightarrow$
$\Phi$ is hermitian-preserving. $\Phi$ is said to be completely
bounded if $\|\phi\|_{cb}=\sup_n \|\Phi_n\| <\infty$. $\Phi:
{\mathcal B}(H)\rightarrow{\mathcal B}(K)$ is called an elementary
operator if there are two finite sequences
$\{A_i\}^n_{i=1}\subset{\mathcal B}(H,K)$ and
$\{B_i\}^n_{i=1}\subset{\mathcal B}(K,H)$ such that
$\Phi(X)=\sum_{i=1}^n A_iXB_i$ for all $X\in{\mathcal B}(H)$; $\Phi:
{\mathcal B}(H)\rightarrow{\mathcal B}(K)$ is called a generalized
elementary operator if there exists sequences $\{A_i\}$ and
$\{B_i\}$
 satisfying $\|\sum_iA_iA_i^\dagger\|\|\sum_iB^\dagger_iB_i\|<\infty$ such that
$\Phi(X)=\sum_iA_iXB_i$ for all $X$.  Obviously, the generalized
elementary operators are completely bounded and normal.

\section{Characterizing   positive completely bounded normal  maps}

In this section we give a characterization of positive completely
bounded normal linear maps from ${\mathcal B}(H)$ into ${\mathcal
B}(K)$. To do this, we need a  lemma.

{\bf Lemma 2.1.} {\it Let $H$, $K$ be separable complex Hilbert
spaces and $\Phi:{\mathcal B}(H)\rightarrow{\mathcal B}(K)$ be a
linear map. Then $\Phi$ is normal and completely bounded if and only
if $\Phi$ is a generalized elementary operator.}

{\bf Proof.} We need only check the ``only if'' part. Assume that
the linear map $\Phi :{\mathcal B}(H)\rightarrow{\mathcal B}(K)$  is
completely bounded and normal. It follows that, $\Phi
=\Phi_1-\Phi_2+i(\Phi_3-\Phi_4)$ with $\Phi_i$  normal and
completely positive by Wittstock's decomposition theorem (ref.
\cite{P}).   As $H$ and $K$ are separable, by Stinespring's Theorem
(ref. \cite{P, St}) and the structural theorem of normal
$*$-homomorphisms of ${\mathcal B}(H)$ (ref. \cite[pp.61]{Dix}), for
each $k=1,2,3,4$, there exist a countable cardinal number $J_k$, an
operator $U_k\in{\mathcal B}(H^{(J_k)}, K)$ such that
$\Phi_k(X)=U_kX^{(J_k)}U_k^\dagger$, where $H^{(J_k)}$ (resp.
$X^{(J_k)}$) is the direct sum of $J_k$-copies of $H$ (resp. of
$X$). Therefore, there are sequences of operators $\{A_i\}_{i\leq
J_1}, \{B_j\}_{j\leq J_2}, \{C_s\}_{s\leq J_3},\{D_t\}_{t\leq
J_4}\subset {\mathcal B}(H,K)$, such that $$U_1=
(\begin{array}{lllll} A_1&A_2&\cdots &A_i&\cdots
\end{array})$$
 $$U_2=(\begin{array}{lllll} B_1&B_2&\cdots &B_j&\cdots \end{array}),$$
 $$U_3= (\begin{array}{lllll} C_1&C_2&\cdots &C_s&\cdots \end{array}),$$
 $$U_4= (\begin{array}{lllll} D_1&D_2&\cdots &D_t&\cdots \end{array})$$ and
 $$\Phi(X)=\sum_{i\leq J_1}A_iXA_i^\dagger-\sum_{j\leq J_2}B_jXB_j^\dagger +i\sum_{s\leq J_3}C_sXC_s^\dagger-
 i\sum_{t\leq J_4}D_tXD_t^\dagger$$ for every $X\in{\mathcal B}(H)$. Now it
 is clear that
$$ \|\sum_{i\leq J_1}A_iA_i^\dagger+\sum_{j\leq J_2}B_iB_i^\dagger+\sum_{s\leq J_3}C_sC_s^{\dagger}
+\sum_{\leq J_4}tD_tD_t^\dagger\|\leq
\sum_{k=1}^4\|U_k\|^2<\infty,$$ and so, $\Phi$ is a generalized
elementary operator. \hfill$\square$

By Lemma 2.1, the question of characterizing positive completely
bounded normal linear maps between ${\mathcal B}(H)$ and ${\mathcal
B}(K)$ is equivalent to the question of  characterizing positive
generalized elementary operators.

As a special class of generalized elementary operators, the global
structures of hermitian-preserving and completely positive
elementary operators are quite clear. In fact, for generalized
elementary operators, by the proof of Lemma 2.1, we have the
following result.

\textbf{Corollary 2.2.} {\it Let $H,K$ be Hilbert spaces and $\Phi$
be a generalized elementary operator from $\mathcal B(H)$ into
$\mathcal B(K)$. Then}

(i) {\it $\Phi$ is hermitian-preserving if and only if there are
sequences  $\{A_i\}, \{C_j\}\subset {\mathcal B}(H,K)$ with
$\parallel \sum _{i=1}^{\infty }A_{i}A_{i}^{\dagger}\parallel
<\infty $ and $\parallel \sum _{j=1}^{\infty
}C_{j}C_{j}^{\dagger}\parallel <\infty $ such that
$$
\Phi(X)=\sum _{i=1}^{\infty }A_{i}XA_{i}^{\dagger}-\sum
_{j=1}^{\infty }C_{j}XC_{j}^{\dagger}
$$
for every $X\in {\mathcal B}(H)$;}

(ii) {\it $\Phi$ is completely positive  if and only if there exists
a sequence  $\{A_i\}\subset {\mathcal B}(H,K)$ with $\parallel \sum
_{i=1}^{\infty }A_{i}A_{i}^{\dagger}\parallel <\infty $ such that
$$
\Phi(X)=\sum _{i=1}^{\infty }A_{i}XA_{i}^{\dagger}
$$
for every $X\in {\mathcal B}(H)$.}

If both $H$ and $K$ are finite-dimensional, Theorem 2.1(i) and (ii)
were established by DePillis \cite{De} and Choi \cite{C1},
respectively. For the elementary operator case, see \cite{H1} and
\cite{M2}.

 For a sequence ${\bf A}=(\begin{array}{ccccc} A_1& A_2& \cdots
 &A_i&\cdots \end{array})$, we will denote by ${\bf A}^T$ the formal
 transpose of $\bf A$ and ${\bf A}^\dagger$ the usual adjoint operator of $\bf A$, that is
 $$ {\bf A}^T=\left(\begin{array}{c} A_1\\A_2\\ \vdots
 \\A_i\\\vdots\end{array}\right)\quad{\rm and}\quad  {\bf A}^\dagger=\left(\begin{array}{c} A_1^\dagger\\A_2^\dagger\\ \vdots
 \\A_i^\dagger\\\vdots\end{array}\right).
$$
 We will also denote by ${\mathcal B}_1(H, K)$  the closed unit ball of
${\mathcal B}(H,K)$.

The next lemma is the key lemma which is a generalization of
\cite[Lemma 2.2]{H4}, where more conditions $\parallel \sum
_{i=1}^{\infty }A_{i}^{\dagger}A_{i}\parallel <\infty $ and
$\parallel \sum _{j=1}^{\infty }C_{j}^{\dagger}C_{j}\parallel
<\infty $ are assumed. Note that, the conditions $\parallel\sum
_{i=1}^{\infty }A_{i}A_{i}^{\dagger}\parallel <\infty $ and
$\parallel \sum _{i=1}^{\infty }A_{i}^{\dagger}A_{i}\parallel
<\infty $  are not equivalent in general. For instance, let
$H=\oplus _{i=1}^\infty H_i$ with each $H_i$ is of infinite
dimension. Let $V_i\in{\mathcal B}(H)$ be the isometry with range
$H_i$. Then $V_i^\dagger V_i=I$ and $V_iV_i^\dagger =P_i$, where
$P_i$ is the projection from $H$ onto $H_i$. Thus
$\|\sum_{i=1}^\infty V_iV_i^\dagger \|=\|\sum_{i=1}^\infty P_i
\|=\|I\|=1$ as $P_iP_j=0$ whenever $i\not= j$, but
$\|\sum_{i=1}^\infty V_i^\dagger V_i\|=\infty$.

{\bf Lemma 2.3.} {\it Let $\{A_{i}\}_{i=1}^{\infty }$ and
$\{C_{j}\}_{j=1}^{\infty }\subset {\mathcal B}(H, K)$ with
$\parallel\sum _{i=1}^{\infty }A_{i}A_{i}^{\dagger}\parallel <\infty
$ and $\parallel\sum _{j=1}^{\infty }C_{j}C_{j}^{\dagger}\parallel
<\infty $. Then the following statements are equivalent:}

(i) {\it $\sum _{i=1}^\infty A_iPA_i^\dagger\geq \sum_{j=1}^\infty
C_jPC_j^\dagger$ for all positive operators $P\in{\mathcal B}(H)$.}

(ii) {\it $\sum _{i=1}^\infty A_iPA_i^\dagger\geq \sum_{j=1}^\infty
C_jPC_j^\dagger$ for all rank-one projections $P\in{\mathcal
B}(H)$.}

(iii) {\it There exists a map $\Omega : H\rightarrow{\mathcal
B}_1(l_2)$ such that
$$ {\bf C}^T|\psi\rangle=\Omega(|\psi\rangle){\bf A}^T|\psi\rangle \quad\mbox{for every }\ |\psi\rangle\in
H.$$}

{\bf Proof.} (i)$\Rightarrow$(ii)$\Rightarrow$(iii) were done in the
proof of \cite[Lemma 2.2]{H4}.

(iii)$\Rightarrow$(ii). Assume (iii). For any unit vector
$|\psi\rangle\in H$, denote $P=|\psi\rangle \langle\psi|$ and the
contractive matrix $\Omega (|\psi\rangle )=\Omega= (\omega_{ij})$.
As $ {\bf C}^T|\psi\rangle=\Omega(|\psi\rangle){\bf
A}^T|\psi\rangle$, we have $C_i|\psi\rangle =\sum_{j=1}^\infty
\omega_{ij}A_j |\psi\rangle$ for each $i$. Thus,
$$\begin{array}{rl} {\bf C}P=&(C_1P, C_2P,\ldots , C_jP,\ldots )\\=&(\sum_{j=1}^\infty
\omega_{1j}A_jP, \sum_{j=1}^\infty \omega_{2j}A_jP, \ldots ,
\sum_{j=1}^\infty \omega_{ij}A_jP, \ldots )\\ =&(A_1P, A_2P, \ldots
, A_jP , \ldots )\Omega ^T \\ =&(A_1P, A_2P, \ldots , A_jP , \ldots
)(w_{ij}I) ^T ={\bf A}P(w_{ij}I) ^T.\end{array}$$ It follows that
$$\sum_{i=1}^\infty C_iPC_i^\dagger ={\bf C}P{\bf C}^\dagger={\bf
A}P(\omega_{ij}I)^T((\omega_{ij}I)^T)^\dagger P{\bf A}^\dagger \leq
{\bf A}P{\bf A}^\dagger= \sum_{j=1}^\infty A_jPA_j^\dagger $$
because of $0\leq (\omega_{ij}I)^T((\omega_{ij}I)^T)^\dagger\leq I$.

(ii)$\Rightarrow$(i). Let  $\Delta (X)=\sum_{j=1}^\infty
A_jXA_j^\dagger - \sum_{i=1}^\infty C_iXC_i^\dagger={\bf
A}X^{(\infty )} {\bf A}^\dagger -{\bf C}X^{(\infty )} {\bf
C}^\dagger $ for each $X\in{\mathcal B}(H)$. Since $\|{\bf
A}\|=\|{\bf A}{\bf A}^\dagger\|^{\frac{1}{2}}=(\|\sum_{j=1}^\infty
A_jA_j^\dagger \|)^{\frac{1}{2}}<\infty$ and $\|{\bf C}\|=\|{\bf
C}{\bf C}^\dagger\|^{\frac{1}{2}}=(\|\sum_{i=1}^\infty
C_iC_i^\dagger \|)^{\frac{1}{2}}<\infty$, we see that $\Delta$ is
normal. The condition (ii) implies that $\Delta (P)$ is positive for
every finite rank positive operator $P$. For any positive operator
$X\in {\mathcal B}(H)$, by spectral theorem, there exists a net
$P_\lambda$ of finite-rank positive operators such that $\|P_\lambda
\|\leq\|X\|$ and ${\rm wk-}\lim_\lambda P_\lambda =X$.  Hence
$\Delta (X)={\rm wk-}\lim_\lambda\Delta(P_\lambda )$ is positive and
(i) is true. \hfill$\square$

The next Lemma  is   a generalization of the main result
\cite[Theorem 2.4]{H4}.

 \textbf{Lemma 2.4.} {\it Let $H$, $K$ be  complex
 Hilbert spaces and
$\{A_{i}\}_{i=1}^{\infty },\{C_{j}\}_{j=1}^{\infty }\subset
{\mathcal B}(H, K)$ with $\parallel \sum _{i=1}^{\infty
}A_{i}A_{i}^{\dagger}\parallel <\infty $ and $\parallel \sum
_{j=1}^{\infty }C_{j}C_{j}^{\dagger}\parallel <\infty $. Let $\Phi
:{\mathcal B}(H)\rightarrow {\mathcal B}(K)$ be a linear map defined
by
$$
\Phi(X)=\sum _{i=1}^{\infty }A_{i}XA_{i}^{\dagger}-\sum
_{j=1}^{\infty }C_{j}XC_{j}^{\dagger}
$$
for every $X\in {\mathcal B}(H)$. Then }

(i) {\it $\Phi$ is positive if and only if there exists a map
$\Omega :|\psi\rangle\in H\mapsto \Omega(|\psi\rangle)=(\omega
_{ji}(|\psi\rangle))_{j,i}\in {\mathcal B}_{1}(l_{2})$ such that
$$
{\bf C}^{T}|\psi\rangle=\Omega(|\psi\rangle){\bf A}^{T}|\psi\rangle
$$
for every $|\psi\rangle\in H$.}

(ii) {\it $\Phi$ is completely positive if and only if there exists
a contractive matrix $\Omega =(\omega _{ji})_{j,i}\in {\mathcal
B}(l_{2})$ such that
$$
{\bf C}^{T}=\Omega {\bf A}^{T},
$$
and in turn, if and only if there exists a sequence
$\{D_{i}\}_{i=1}^{\infty }\subset {\mathcal B}(H,K)$ such that
$$
\Phi(X)=\sum _{i=1}^{\infty }D_{i}XD_{i}^{\dagger}.
$$
holds for all $X\in {\mathcal B}(H)$.

Here ${\bf A}=(A_1, A_2, \ldots , A_n, \ldots )$ and ${\bf C}=(C_1,
C_2, \ldots , C_n, \ldots )$.}

{\bf Proof.} By Lemma 2.3, \cite[Theorem 2.4]{H4} and its proof, we
know that the lemma holds   except the conclusion that $\Phi$ is
completely positive if and only if there exists a sequence
$\{D_{i}\}_{i=1}^{\infty }\subset {\mathcal B}(H,K)$ such that
$$
\Phi(X)=\sum _{i=1}^{\infty }D_{i}XD_{i}^{\dagger}
$$
for every $X\in {\mathcal B}(H)$. But this is true by Corollary 2.2.
\hfill$\square$

Combine Lemma 2.1 and Lemma 2.4, one gets the main result of this
section immediately.

{\bf Theorem 2.5.} {\it  Let $H$, $K$ be separable complex
 Hilbert spaces and $\Phi :{\mathcal B}(H)\rightarrow{\mathcal
 B}(K)$ be a completely bounded normal linear map. Then }

 (1) {\it $\Phi$ is
 positive if and only if there exist $\{A_{i}\}_{i=1}^{\infty },\{C_{j}\}_{j=1}^{\infty }\subset
{\mathcal B}(H, K)$ with $\parallel \sum _{i=1}^{\infty
}A_{i}A_{i}^{\dagger}\parallel <\infty $ and $\parallel \sum
_{j=1}^{\infty }C_{j}C_{j}^{\dagger}\parallel <\infty $, and a map
$\Omega :|\psi\rangle\in H\mapsto \Omega(|\psi\rangle)=(\omega
_{ji}(|\psi\rangle))_{j,i}\in {\mathcal B}_{1}(l_{2})$ satisfying
$$
{\bf C}^{T}|\psi\rangle=\Omega(|\psi\rangle){\bf A}^{T}|\psi\rangle
$$
for every $|\psi\rangle\in H$, such that $$ \Phi(X)=\sum
_{i=1}^{\infty }A_{i}XA_{i}^{\dagger}-\sum _{j=1}^{\infty
}C_{j}XC_{j}^{\dagger}
$$
holds for every $X\in {\mathcal B}(H)$.}

(2) {\it $\Phi$ is completely positive if and only if there exists a
sequence $\{D_{i}\}_{i=1}^{\infty }\subset {\mathcal B}(H,K)$ with
$\parallel \sum _{i=1}^{\infty }D_{i}D_{i}^{\dagger}\parallel
<\infty $ such that
$$
\Phi(X)=\sum _{i=1}^{\infty }D_{i}XD_{i}^{\dagger}.
$$
holds for all $X\in {\mathcal B}(H)$.

Here ${\bf A}=(A_1, A_2, \ldots , A_n, \ldots )$ and ${\bf C}=(C_1,
C_2, \ldots , C_n, \ldots )$.}

What does Theorem 2.5 mean? To understand Theorem 2.5 better, let us
recall some notions from \cite{H4}.
 Let $l$, $k\in\mathbb{N}$ (the set
of all natural numbers),  and let $A_{1},\cdots, A_{k}$, and
$C_{1},\cdots, C_{l}\in {\mathcal B}(H$, $K$). If, for each
$|\psi\rangle\in H$, there exists an $l\times k$ complex matrix
$(\alpha _{ij}(|\psi\rangle))$ (depending on $|\psi\rangle$) such
that
$$
C_{i}|\psi\rangle=\sum _{j=1}^{k}\alpha
_{ij}(|\psi\rangle)|A_{j}\psi\rangle,\qquad i=1,2,\cdots ,l,
$$
we say that $(C_{1},\cdots ,C_{l})$ is a locally linear combination
of $(A_{1},\cdots ,A_{k})$, $(\alpha_{ij}(|\psi\rangle))$ is called
a {\it local coefficient matrix} at $|\psi\rangle$.  Furthermore, if
 a local coefficient matrix
$(\alpha_{ij}(|\psi\rangle))$
  can be chosen for every $|\psi\rangle\in H^{(n)}$ so that
the operator norm $\|(\alpha _{ij}(|\psi\rangle))\|\leq 1$, we say
that $(C_{1},\cdots ,C_{l})$ is a {\it contractive locally linear
combination} of $(A_{1},\cdots ,A_{k})$; if there is a matrix
$(\alpha_{ij})$ with $\|(\alpha _{ij}\|\leq 1$ such that $C_{i}=\sum
_{j=1}^{k}\alpha _{ij}A_{j}$ for all $i$, we say that $(C_{1},\cdots
,C_{l})$ is a {\it contractive linear combination} of $(A_{1},\cdots
,A_{k})$ with coefficient matrix $(\alpha _{ij})$.  Sometimes we
also write $\{A_i\}_{i=1}^k$ for $(A_{1},\cdots ,A_{k})$. These
notions can be generalized to the case that there are infinite many
$A_k$s or $C_k$s. For instance, if, for every $|\psi\rangle\in H$,
there are scalars $\alpha_k(|\psi\rangle )$ such that
$C|\psi\rangle=\sum_{k=1}^\infty\alpha_k(|\psi\rangle)
A_k|\psi\rangle$ and $\sum_{k=1}^\infty
|\alpha_k(|\psi\rangle)|^2\leq 1$, we will say that $C$ is a {\it
generalized contractive locally linear combination} of
$\{A_k\}_{k=1}^\infty$.

Thus Theorem 2.5 may be restated as follows: A  completely bounded
normal linear map $\Phi :{\mathcal B}(H)\rightarrow{\mathcal
 B}(K)$ is positive but not completely positive (briefly, NCP) if and only if it has the form $ \Phi(X)=\sum
_{i=1}^{\infty }A_{i}XA_{i}^{\dagger}-\sum _{j=1}^{\infty
}C_{j}XC_{j}^{\dagger}$ for all $X$, where $\{C_j\}$ is a
generalized contractive locally linear combination of $\{A_i\}$ but
$\{C_j\}$ is not a generalized contractive  linear combination of
$\{A_i\}$. This characterization is much helpful in some sense  to
understand the differences of completely positive normal linear
maps, positive completely bounded normal linear maps and hermitian
completely bounded normal linear maps.

By Theorem 2.5, one gets immediately   a global structure theorem
for positive elementary operators in terms of local linear
combination that was established in \cite{H4}. For ${\mathcal
L}\subset{\mathcal B}(H,K)$, we'll denote by $[{\mathcal L}]$ the
linear span of ${\mathcal L}$.

\textbf{Corollary 2.6.} {\it Let $\Phi =\sum _{i=1}^{n}A_{i}(\cdot
)B_{i}$ be an elementary operator from ${\mathcal B}(H)$ into
${\mathcal B}(K)$. Then $\Phi $ is positive if and only if there
exist $C_{1},\cdots ,C_{k}$ and $D_{1},\cdots ,D_{l}$ in
$[A_{1},\cdots ,A_{n}]$ with $k+l\leq n$ such that $(D_{1},\cdots
,D_{l})$ is an contractive locally linear combination of
$(C_{1},\cdots ,C_{k})$ and
$$
\Phi =\sum _{i=1}^{k}C_{i}(\cdot )C_{i}^{\dagger}-\sum
_{j=1}^{l}D_{j}(\cdot )D_{j}^{\dagger}.\eqno(2.1)
$$
Furthermore, $\Phi$ in Eq.(2.1) is completely positive if and only
if $(D_{1},\cdots ,D_{l})$ is a linear combination of $(C_{1},\cdots
,C_{k})$ with a contractive coefficient matrix, and in turn, if and
only if there exist $E_1, E_2, \ldots , E_r$ with $r\leq k$ such
that
$$ \Phi=\sum_{i=1}^r E_i(\cdot )E_i^\dagger.$$}

Since every linear map between matrix algebras is an elementary
operator, by Corollary 2.6 we get a characterization of positive
maps that is not CP for finite dimensional case.

{\bf Corollary 2.7.}  {\it Let $H$ and $K$ be finite dimensional
complex Hilbert spaces and let $\Phi :{\mathcal
B}(H)\rightarrow{\mathcal B}(K)$ be a linear map. Then $\Phi$ is
positive but not completely positive if and only if there exist
$C_{1},\cdots ,C_{k},D_{1},\cdots ,D_{l}\in{\mathcal B}(H,K)$ such
that  $\Phi
(X)=\sum_{i=1}^kC_iXC_i^\dagger-\sum_{j=1}^lD_jXD_j^\dagger$ for all
$X\in{\mathcal B}(H)$, and $\{D_j\}_{j=1}^l$ is a contractive
locally linear combination  but not a contractive linear combination
of $\{C_i\}_{i=1}^k$.}

It is interesting to observe from the discussion above that, for
elementary operators, the question when positivity ensures complete
positivity may be reduced to the question when contractive locally
linear combination implies linear combination. This connection
allows us to look more deeply into the relationship and the
difference between positivity and complete positivity, and obtain
some simple criteria to check whether a positive elementary operator
is completely positive or not. This is important especially when we
construct positive maps and apply them to recognize entanglement.

If ${\mathcal L}\subset{\mathcal B}(H,K)$, we will denote by
${\mathcal L}_F$ the subset of all finite-rank operators in
${\mathcal L}$.

The Corollaries 2.8 and 2.9  below can be found in \cite{H4}. We
list them here for completeness and for reader's convenience.

{\bf Corollary 2.8.}  {\it  Assume that $\Phi =\sum
_{i=1}^{k}A_{i}(\cdot )A^*_{i}-\sum _{j=1}^{l}B_{j}(\cdot
)B^*_{j}:{\mathcal B}(H)\rightarrow {\mathcal B}(K)$ is a positive
elementary operator. If any one of the following conditions holds,
then $\Phi $ is completely positive:}

(i) {\it $k\leq 2$.}

(ii) {\it $\dim[A_{1},\cdots ,A_{k}]_{F}\leq 2$.}

(iii) {\it  There exists a vector $|\psi\rangle\in H$  such that
$\{|A_{i}\psi\rangle\}_{i=1}^{k}$  is linearly independent.}

(iv) {\it $\Phi $ is  $[\frac{k+1}{2}]$-positive, where $[t]$ stands
for the integer part of real number $t$.}

{\bf Corollary 2.9.}  {\it  Assume that $\Phi =\sum
_{i=1}^{k}A_{i}(\cdot )A^*_{i}-\sum _{j=1}^{l}B_{j}(\cdot
)B^*_{j}:{\mathcal B}(H)\rightarrow {\mathcal B}(K)$ is a positive
elementary operator. If  $\Phi $ is not completely positive, then }

(i) {\it $k\geq 3$,}

(ii) {\it $\dim[A_{1},\cdots ,A_{k}]_{F}\geq 3$,}

(iii) {\it $B_j$, $j=1,2,\ldots ,l$, is a finite-rank perturbation
of some combination of $\{A_{i}\}_{i=1}^{k}$.}

(iv) {\it $\Phi _{[\frac{k+1}{2}]}$ is not positive.}

{\bf Corollary 2.10.} {\it Assume that $\Phi =\sum
_{i=1}^{k}A_{i}(\cdot )A^*_{i}-\sum _{j=1}^{l}B_{j}(\cdot
)B^*_{j}:{\mathcal B}(H)\rightarrow {\mathcal B}(K)$ is an
elementary operator. If  there exists some $j$ such that $B_j$ is
not a contractive linear combination of $\{A_{i}\}_{i=1}^{k}$, then
$\Phi $ is not completely positive.}

The following result reveals that the non-complete positivity of a
positive elementary operator is essentially determined by its
behavior on finite-dimensional subspaces.  So, to construct a NCP
positive elementary operator, it is enough to consider the question
in finite-dimensional cases.

{\bf Theorem 2.11.} {\it Assume that $\Phi :{\mathcal
B}(H)\rightarrow {\mathcal B}(K)$ is a positive elementary operator.
Then $\Phi$ is NCP if and only if there exist finite-rank
projections $P$ and $Q$ acting on $H$ and $K$, respectively, such
that the positive  elementary operator $\Delta : {\mathcal
B}(PH)\rightarrow{\mathcal B}(QK)$ defined by $\Delta
(X)=Q\Phi(PXP)Q|_{QK}$ is non-completely positive. In addition, $P$
and $Q$ may be taken so that $\Delta':{\mathcal B}(\ker
P)\rightarrow{\mathcal B}(\ker Q) $ defined by $\Delta'
(Y)=(I-Q)\Phi(((I-P)Y(I-P))(I-Q)|_{\ker Q}$ is completely positive.}

{\bf Proof.} Clearly, if $\Phi :{\mathcal B}(H)\rightarrow {\mathcal
B}(K)$ is a positive linear map and $P\in{\mathcal B}(H)$,
$Q\in{\mathcal B}(K)$ are projections, then $\Delta : {\mathcal
B}(PH)\rightarrow{\mathcal B}(QK)$ defined by $\Delta
(X)=Q\Phi(PXP)Q$ is positive and $\Delta$ is NCP
 implies that $\Phi$ is NCP.

Assume that $\Phi$ is a positive elementary operator, writing $\Phi
=\sum _{i=1}^{k}A_{i}(\cdot )A^*_{i}-\sum _{j=1}^{l}B_{j}(\cdot
)B^*_{j}$ with $\{A_1, \ldots , A_k, B_1, \ldots , B_l\}$ linearly
independent. By Corollary 2.9 (ii)-(iii), if $\Phi$ is NCP, then the
linear subspace spanned by $\{A_i\}_{i=1}^k$ has many finite rank
operators and there exists $C_j\in[A_1,A_2,\ldots , A_k]$ and finite
rank operators $F_j\not\in [A_1,\ldots , A_k]$ such that
$B_j=C_j+F_j$. Let $P_0$ be the projection with range the finite
dimensional linear subspace spanned by all the ranges of $\{
E^\dagger : E\in [A_1, \ldots , A_k]_{\mathcal F}\}$ and the ranges
of $\{F_j^\dagger\}_{j=1}^l$; and $Q_0$ the projection with range
the finite dimensional linear subspace spanned by all the ranges of
$\{ E : E\in [A_1, \ldots , A_k]_{\mathcal F}\}$ and the ranges of
$\{ F_j\}_{j=1}^l$. It is easily checked that there exist some
finite rank projections $P\geq P_0$ and $Q\geq Q_0$ such that
$QB_jP\not\in[QA_1P, \ldots , QA_kP]$ since $B_j\not\in [A_1,\ldots
,A_k]$. Pick such $P$ and $Q$. Let $S_i=QA_i|_{PH}$, $i=1,2,\ldots
,k$, and $T_j=QB_j|_{PH}$, $j=1,2,\ldots ,l$. Let $\Delta :{\mathcal
B}(PH)\rightarrow{\mathcal B}(QK)$ be the map defined by $\Delta
(X)= \sum _{i=1}^{k}S_{i}XS^*_{i}-\sum
_{j=1}^{l}T_{j}XT^*_{j}=Q\Phi(PXP)Q|_{QK}$. Then $\Delta$ is
positive. By the choice of $P$ and $Q$, $T_j$ is not in $[S_1,
\ldots , S_k]$
  for some $j$. Hence, $\Delta$ is
not completely positive by Corollary 2.9. Since $[(I-Q)A_1(I-P),
\ldots, (I-Q)A_k(I-P)]_{\mathcal F}=\{0\}$, by Corollary, $\Delta'$
is completely positive. \hfill$\Box$

To conclude this section, we give a simple example illustrating that
how to use the results in this section to judge whether or not a map
is positive, completely positive.

{\bf Example 2.12.} Assume that $\dim H=n$ and
$\{|i\rangle\}_{i=1}^n$ is an orthonormal basis. Denote
$E_{ij}=|i\rangle\langle j|$. For a given positive number $t$, let
$\Delta_t :{\mathcal B}(H)\rightarrow{\mathcal B}(H)$ be a linear
map defined by
$$\Delta_t(X)=t\sum_{i=1}^n E_{ii}XE_{ii}-X$$ for any $X\in{\mathcal
B}(H)$. Then $\Delta_t$ is positive if and only if it is completely
positive, and in turn, if and only  if $t\geq n$.

In fact, let $A_i=\sqrt{t}E_{ii}$, then $\Delta_t(X)=\sum_{i=1}^n
A_iXA_i^\dagger -IXI^\dagger$. It is clear that $I$ is a linear
combination of $A_1,\cdots ,A_n$, i.e.,
$I=\sum_{i=1}^n\frac{1}{\sqrt{t}}A_i$. Then the sum of the square of
the coefficients is $\sum_i(\frac{1}{\sqrt{t}})^2=\frac{n}{t}$, and
hence $\Delta_t$ is completely positive if and only if $t\geq n$ by
Corollary 2.6. If $t< n$, then it is obvious that $I$ is not a
contractive locally linear combination of $A_1,\cdots , A_n$, and
hence $\Delta_t$ is not positive.

\section{Characterizing  quantum channels for infinite dimensional systems}

It is known that, for finite-dimensional quantum systems, a quantum
channel (operation) $\mathcal E$ is a trace-preserving
(trace-nonincreasing) completely positive linear map between
associated matrix algebras. Thus, by a result due to Choi \cite{C1},
$\mathcal E$ is an elementary operator of the form ${\mathcal
E}(\cdot)=\sum_{i=1}^nA_i(\cdot)A_i^{\dagger}$, where
$\sum_{i=1}^nA_i^\dagger A_i=I$ ($\sum_{i=1}^nA_i^\dagger A_i\leq
I$). Using the discussion in Section 2, one can characterize the
completely bounded linear maps, positive completely bounded linear
maps and completely positive linear maps between the trace-classes.
This allow us to obtain a similar representation of quantum
operations for infinite-dimensional systems. Firstly we recall some
notions. For $A\in{\mathcal B}(H)$, denote $|A|=(A^\dagger
A)^{\frac{1}{2}}$. Recall that the trace class ${\mathcal T}(H)=\{T
: \|T\|_1={\rm Tr}(|T|)<\infty\}$, which is a ideal of ${\mathcal
B}(H)$. Furthermore, ${\mathcal T}(H)$ is a Banach space with the
trace norm $\|\cdot\|_1$. The dual space of ${\mathcal T}(H)$ is
${\mathcal T}(H)^*={\mathcal B}(H)$ and every bounded linear
functional is of the form $T\mapsto{\rm Tr}(AT)$, where
$A\in{\mathcal B}(H)$.

{\bf Lemma 3.1.} {\it Let $H$, $K$ be separable complex Hilbert
spaces and ${\mathcal T}(H)$, ${\mathcal T}(K)$ be the trace classes
on $H$, $K$ respectively. Then, a linear map $\Delta :{\mathcal
T}(H)\rightarrow{\mathcal T}(K)$ is completely bounded if and only
if there exists operator sequences $\{A_i\}_i\subset{\mathcal
B}(H,K)$ and $\{B_i\}_i\subset{\mathcal B}(K,H)$ satisfying
$\parallel \sum _{i}A_{i}^{\dagger} A_{i}\parallel <\infty $,  and
$\parallel \sum _{i}B_{i} B_{i}^{\dagger}\parallel <\infty $ such
that
$$
\Delta(T)=\sum _{i }A_{i}TB_{i}
$$
for all $T\in{\mathcal T}(H)$.}

{\bf Proof.} If $\Delta$ has the form stated in the theorem, it is
obvious that, for any $X\in{\mathcal B}(K)$,
$${\rm Tr}(\sum _{i }A_{i}TB_{i}X)=\sum _{i }{\rm
Tr}(A_{i}TB_{i}X) =\sum _{i }{\rm Tr}(TB_iXA_{i})={\rm Tr}(\sum _{i
}TB_iXA_{i})$$ holds for all $T\in{\mathcal T}(H)$, so
$\Delta^*(X)=\sum _{i }B_iXA_{i}\in{\mathcal B}(H)$. As $\parallel
\sum _{i}A_{i}^{\dagger}A_{i}\parallel <\infty $,  and $\parallel
\sum _{i}B_{i}B_{i}^{\dagger}\parallel <\infty $, $\Delta ^*$ is
completely bounded with $\|\Delta^*\|_{\rm cb}\leq \parallel (\sum
_{i}A_{i}^{\dagger}A_{i})^{\frac{1}{2}}\parallel\cdot\parallel(\sum
_{i}B_{i}B_{i}^{\dagger})^{\frac{1}{2}}\parallel.$ But
$\|\Delta_n\|=\|\Delta_n^*\|$ (ref. \cite[Proposition 3.2.2]{ER}),
so, $\Delta$ is completely bounded.

Conversely, assume that $\Delta :{\mathcal T}(H)\rightarrow{\mathcal
T}(K)$ is a completely bounded linear map; then $\Delta^*:{\mathcal
B}(K)\rightarrow{\mathcal B}(H)$ is a completely bounded normal
linear map. By Lemma 2.1, $\Delta ^*$ is a generalized elementary
operator. So there exists operator sequences
$\{A_i\}_i\subset{\mathcal B}(H,K)$ and $\{B_i\}_i\subset{\mathcal
B}(K,H)$ satisfying $\parallel \sum _{i}A_{i}^{\dagger}
A_{i}\parallel <\infty $,  and $\parallel \sum _{i}B_{i}
B_{i}^{\dagger}\parallel <\infty $ such that $\Delta ^*(X)=\sum_{i}
B_iXA_i$ holds for all $X\in {\mathcal B}(K,H)$. Now, it is clear
that $\Delta (T)=\sum_{i}A_iT B_i$ holds for all $T\in {\mathcal
B}(K,H)$, completing the proof.\hfill$\square$

By Lemma 3.1 and Theorem 2.5 the following results are immediate.

{\bf Theorem 3.2.} {\it Let $H$, $K$ be separable complex Hilbert
spaces and ${\mathcal T}(H)$, ${\mathcal T}(K)$ be the trace classes
on $H$, $K$ respectively. Let  $\Delta :{\mathcal
T}(H)\rightarrow{\mathcal T}(K)$ be a linear map. Then }

(i) {\it $ \Delta$ is positive and completely bounded if and only if
there exists operator sequences $\{A_i\}_i\subset{\mathcal B}(H,K)$
and $\{B_i\}_i\subset{\mathcal B}(H,K)$ with $\parallel \sum
_{i}A_{i}^{\dagger}A_{i}\parallel <\infty $  and $\parallel \sum
_{i}B_{i}^{\dagger}B_{i}\parallel <\infty $, and a map $\Omega
:H\rightarrow{\mathcal B}_1(l_2)$  such that ${\bf
B}^\dagger|\psi\rangle =\Omega(|\psi\rangle ){\bf
A}^\dagger|\psi\rangle$ for every $|\psi\rangle \in H$ and
$$
\Delta(T)=\sum _{i }A_{i}TA_{i}^\dagger-\sum _{i
}B_{j}TB_{j}^\dagger
$$
for all $T\in{\mathcal T}(H)$.}

(ii) {\it $ \Delta$ is  completely positive if and only if there
exists operator sequences $\{A_i\}_i\subset{\mathcal B}(H,K)$  with
$\parallel \sum _{i}A_{i}^{\dagger}A_{i}\parallel <\infty $    such
that
$$
\Delta(T)=\sum _{i }A_{i}TA_{i}^\dagger
$$
for all $T\in{\mathcal T}(H)$.}\\

{\bf Corollary 3.3.} {\it Every quantum channel (operation)
$\mathcal E$ between two infinite-dimensional systems respectively
associated with Hilbert spaces $H$ and $K$  has the form
$$ {\mathcal E}(\rho)=\sum_{i=1}^\infty M_i\rho M_i^\dagger,
$$
where $\{M_i\}\subset{\mathcal B}(H,K)$ satisfies that
$\sum_{i=1}^\infty M_i^\dagger M_i=I_H$ ($\sum_{i=1}^\infty
M_i^\dagger M_i\leq I_H$).}

\section{Elementary operator criterion of separability}

Using the characterization of positive maps that are NCP in Section
2, we can get some criteria of entanglement of quantum states based
on the positive map criterion. These will help us to deduce a
necessary and sufficient criterion of separability of states.

The following necessary and sufficient condition for a state on
finite dimensional spaces to be entangled is an immediate
consequence of Corollary 2.7 and Horodeckis' Theorem.

{\bf Theorem 4.1.} {\it Let $H$ and $K$ be finite dimensional
complex Hilbert spaces and $\rho$ be a state acting on $H\otimes K$.
Then $\rho$ is an entangled state if and only if there exists a
linear map of the form $\Phi(\cdot )= \sum _{i=1}^{k}C_{i}(\cdot
)C_{i}^{\dagger}-\sum _{j=1}^{l}D_{j}(\cdot
)D_{j}^{\dagger}:{\mathcal B}(H)\rightarrow {\mathcal B}(K)$ with
 $\{ D_1,\ldots , D_l\}$  a contractive locally linear
combination of $\{C_1, \ldots ,C_k\}$, such that the operator
$(\Phi\otimes I)\rho $ is not positive.}

We will show below that this result is also true for infinite
dimensional case. Before doing this, we write directly from Theorem
2.5 and Corollary 2.6 two sufficient criteria of entanglement of
states for infinite dimensional systems.

{\bf Proposition 4.2.} {\it Let $H, K$ be complex Hilbert spaces
and $\rho$ be a state on $H\otimes K$. Then $\rho$ is entangled if
there exists an elementary operator  of the form $\Phi(\cdot )= \sum
_{i=1}^{k}C_{i}(\cdot )C_{i}^{\dagger}-\sum _{j=1}^{l}D_{j}(\cdot
)D_{j}^{\dagger}:{\mathcal B}(H)\rightarrow {\mathcal B}(K)$, where
 $\{ D_1,\ldots , D_l\}$ is a
contractive locally linear combination but not a contractive linear
combination of $\{C_1, \ldots ,C_k\}$, such that the operator
$(\Phi\otimes I)\rho $ is not positive.}

More generally,  we have

{\bf Proposition 4.3.} {\it Let $H, K$ be complex Hilbert spaces and
$\rho$ be a state on $H\otimes K$. Then $\rho$ is an entangled state
if there exists a generalized elementary operator $\Phi$ defined by
$$
\Phi(X)=\sum _{i}A_{i}XA_{i}^{\dagger}-\sum
_{j}C_{j}XC_{j}^{\dagger}
$$
for every $X\in {\mathcal B}(H)$, where $\parallel \sum
_{i}A_{i}A_{i}^{\dagger}\parallel <\infty $ and $\parallel \sum _{j
}C_{j}C_{j}^{\dagger}\parallel <\infty $, $\{C_{j}\}_{j}$ is a
generalized contractive locally linear combination but not a
generalized contractive linear combination of $\{A_{i}\}_{i}$, such
that $(\Phi\otimes I)\rho$ is not positive.}

Propositions 4.2 and 4.3 only provide  sufficient conditions for a
state to be entangled and are not easily applied practically. In
fact, these conditions are also necessary, and thus we obtain a
necessary and sufficient criterion for entanglement which we will
call the elementary operator criterion. Much better can be reached.
Note that an elementary operator $\Phi$ is of finite rank if and
only if there exist finite rank operators $A_i, B_i$, $i=1,2,\cdots
,k$, such that $\Phi(X) =\sum_{i=1}^k A_iXB_i$ \cite{H5}. We will
prove that every entangled state can be detected by a positive
elementary operator of finite rank.

{\bf Theorem 4.4.} (Elementary operator criterion) {\it  Let $H, K$
be complex Hilbert spaces  and $\rho$ be a state on $H\otimes K$.
Then $\rho$ is entangled if and only if there exists an elementary
operator of the form $\Phi(\cdot )= \sum _{i=1}^{k}C_{i}(\cdot
)C_{i}^{\dagger}-\sum _{j=1}^{l}D_{j}(\cdot
)D_{j}^{\dagger}:{\mathcal B}(H)\rightarrow {\mathcal B}(K)$, where
 all $C_i$s and $D_j$s are of finite rank and $\{ D_1,\ldots , D_l\}$ is a
contractive locally linear combination of $\{C_1, \ldots ,C_k\}$,
such that the operator $(\Phi\otimes I)\rho $ is not positive.}

{\bf Proof.} The ``only if'' part follows from Proposition 4.2. For
``if'' part, assume that the state $\rho$ is inseparable.  Take any
orthonormal bases $\{|i\rangle\}$ and $\{|j\rangle\}$ of $H $ and
$K$, respectively. For any positive integers $s\leq \dim H$ and
$t\leq\dim K$, denote $P_{st}=P_s\otimes Q_t$, where $P_s$ and $Q_t$
are finite rank projections onto the subspaces $H_s$ and $K_t$
spanned by $\{|i\rangle\}_{i=0}^s$ and $\{|j\rangle\}_{j=0}^t$,
respectively. Since $\rho$ is entangled, by \cite[Theorem 2]{SV2},
there exists $(s,t)$ such that $\rho_{st}={\rm Tr}(P_{st}\rho
P_{st})^{-1}P_{st}\rho P_{st} $ is entangled. Regarding $\rho_{st}$
as a state on $H_s\otimes K_t$. As $\dim (H_s\otimes K_t)<\infty$,
by Theorem 4.1, there exists a positive map $\Delta :{\mathcal
B}(H_s)\rightarrow{\mathcal B}(K_t)$ of the form $\Delta(\cdot )=
\sum _{i=1}^{k}A_{i}(\cdot )A_{i}^{\dagger}-\sum
_{j=1}^{l}B_{j}(\cdot )B_{j}^{\dagger}$ with
 $\{ B_1,\ldots , B_l\}$  a contractive locally linear
combination but not a contractive linear combination of $\{A_1,
\ldots ,A_k\}$, such that the operator $(\Delta\otimes Q_t)\rho_{st}
$ is not positive on $K_t\otimes K_t$. Let $\Phi:{\mathcal
B}(H)\rightarrow{\mathcal B}(K)$ be defined by
$\Phi(X)=Q_t\Delta(P_sXP_s)Q_t$. Then $\Phi$ is positive and
$\Phi(X)=\sum _{i=1}^{k}C_{i}(X )C_{i}^{\dagger}-\sum
_{j=1}^{l}D_{j}(X )D_{j}^{\dagger}$, where $C_i=Q_tA_iP_s$ and
$D_j=Q_tB_iP_s$ are of finite rank.

Represent $\rho$ as an operator matrix $\rho=(\eta_{ij})_{i,j}$
according to the bases $\{|i\rangle\}_{i=0}^s$ and
$\{|j\rangle\}_{j=0}^t$, where $\eta_{ij}\in{\mathcal B}(H)$.
Obviously, $$\rho_{st}={\rm Tr}(P_{st}\rho
P_{st})^{-1}\left(\begin{array}{cccc}
P_s\eta_{11}P_s & P_s\eta_{12}P_s &\cdots &P_s\eta_{1t}P_s\\
P_s\eta_{21}P_s & P_s\eta_{22}P_s &\cdots &P_s\eta_{2t}P_s\\
\vdots &\vdots &\ddots &\vdots \\
P_s\eta_{t1}P_s & P_s\eta_{t2}P_s &\cdots &P_s\eta_{tt}P_s \\
\end{array}\right).$$ Thus we have
$$(\Delta\otimes Q_t)\rho_{st}={\rm Tr}(P_{st}\rho
P_{st})^{-1}\left(\begin{array}{cccc}
\Delta(P_s\eta_{11}P_s) & \Delta(P_s\eta_{12}P_s) &\cdots &\Delta(P_s\eta_{1t}P_s)\\
\Delta(P_s\eta_{21}P_s) & \Delta(P_s\eta_{22}P_s) &\cdots &\Delta(P_s\eta_{2t}P_s)\\
\vdots &\vdots &\ddots &\vdots \\
\Delta(P_s\eta_{t1}P_s) & \Delta(P_s\eta_{t2}P_s) &\cdots &\Delta(P_s\eta_{tt}P_s) \\
\end{array}\right)\eqno(4.1)$$ is not positive.
Note that
$\Phi(\eta_{ij})=Q_t\Delta(P_s\eta_{ij}P_s)Q_t=\Delta(P_s\eta_{ij}P_s)$.
So
$$\begin{array}{rl} &(\Phi\otimes I)\rho\\=&\left(\begin{array}{cccc|cc}
\Delta(P_s\eta_{11}P_s) & \Delta(P_s\eta_{12}P_s) &\cdots &\Delta(P_s\eta_{1t}P_s) &\Delta(P_s\eta_{1(t+1)}P_s)&\cdots\\
\Delta(P_s\eta_{21}P_s )& \Delta(P_s\eta_{22}P_s) &\cdots &\Delta(P_s\eta_{2t}P_s )&\Delta(P_s\eta_{2(t+1)}P_s)&\cdots\\
\vdots &\vdots &\ddots &\vdots &\vdots&\ddots\\
\Delta(P_s\eta_{t1}P_s) & \Delta(P_s\eta_{t2}P_s) &\cdots &\Delta(P_s\eta_{tt}P_s) &\Delta(P_s\eta_{t(t+1)}P_s)&\cdots\\
\hline \Delta(P_s\eta_{(t+1)1}P_s) & \Delta(P_s\eta_{(t+1)2}P_s) &\cdots &\Delta(P_s\eta_{(t+1)t}P_s) &\Delta(P_s\eta_{(t+1)(t+1)}P_s)&\cdots\\
\vdots &\vdots &\ddots &\vdots &\vdots&\ddots
\end{array}\right).\end{array}$$
It follows that $(\Phi\otimes I)\rho$ is not positive since it has a
non positive $t\times t$ submatrix (4.1). The proof is
completed.\hfill$\Box$

To sum up, we have proved the following criterion of separability,
which is valid for both finite and infinite dimensional systems,
improves St{\o}mer's theorem \cite{Sto} and is easier to practise by
our characterization of positive elementary operators.

{\bf Theorem 4.5.} (Elementary operator criterion) {\it Let $H$, $K$
be complex Hilbert spaces and $\rho$ be a state acting on $H\otimes
K$. Then the following statements are equivalent.}

(1) {\it $\rho$ is separable;}

(2) {\it $(\Phi\otimes I)\rho \geq 0$ holds for every positive
 elementary operator $\Phi :{\mathcal B}(H)\rightarrow
{\mathcal B}(K)$.}

(3) {\it $(\Phi\otimes I)\rho \geq 0$ holds for every finite-rank
positive
 elementary operator $\Phi :{\mathcal B}(H)\rightarrow
{\mathcal B}(K)$.}

\section{Examples of NCP positive maps and entangled states}

It follows from Theorem 4.4, 4.5 and Theorem 2.11, for both finite
and infinite dimensional systems, it is very important to construct
NCP positive linear maps between matrix algebras since the
non-complete positivity of a positive elementary operator is
essentially determined by its behavior on finite-dimensional
subspaces. In this section we give some concrete examples  of NCP
positive linear maps between matrix algebras   by applying the
results in Section 2, and, according to the elementary operator
criterion, some of them are used to recognize entangled states that
cannot be recognized by the PPT criterion and the realignment
criterion.

 Let $H$ be a complex Hilbert space of
$\dim H=n<\infty$ and let $\{|1\rangle , |2\rangle , \ldots ,
|n\rangle\}$ be an orthonormal basis of $H$. Denote
$E_{ij}=|i\rangle\langle j|$,  $1\leq i,j\leq n$. The well known NCP
positive map on ${\mathcal B}(H)$, that is, the transpose $T\mapsto
T^t$ is an elementary operator
$$T^t=\sum_{i=1}^n E_{ii}TE_{ii} +\sum _{i<j} A_{ij}TA_{ij}^\dagger
-\sum_{i<j}C_{ij}TC_{ij}^\dagger \quad \forall T,$$ where
$A_{ij}=\frac{1}{\sqrt{2}}(E_{ij}+E_{ji})$,
$C_{ij}=\frac{1}{\sqrt{2}}(E_{ij}-E_{ji})$. Another example of well
known NCP positive map is the reduction map, which  has the form
$$T\mapsto {\rm Tr}(T)I-T=\sum_{i\not=j} E_{ij}TE_{ji}+\sum_{i\not=j}G_{ij}AG_{ij}^\dag
-\sum_{i\not=j}F_{ij}AF_{ij}^\dag\quad \forall T,$$ where
$F_{ij}=\frac{1}{\sqrt{2}}(E_{ii}+E_{jj})$ and
$G_{ij}=\frac{1}{\sqrt{2}}(E_{ii}-E_{jj})$.

 Next we give another kind of NCP positive linear maps.

{\bf Proposition 5.1.} {\it   Let $H$ be a complex Hilbert space of
$2\leq \dim H=n<\infty$ and let $\{|1\rangle , |2\rangle , \ldots ,
|n\rangle\}$ be an orthonormal basis of $H$. Denote
$E_{ij}=|i\rangle\langle j|$,  $1\leq i,j\leq n$. Let
$A_k=\sum_{i=1}^n a_{ki}|i\rangle\langle i|$, $k=1, \ldots ,s$ and
$B_l=\sum_{i=1}b_{li}|i\rangle\langle i|$, $l=1,\ldots , t$  with
$t>0$ and $s+t\leq n$. Assume that $\{A_k, B_l: k=1,\ldots , s;
l=1,\ldots , t\} $ is a linearly independent set. Let $\Delta :
{\mathcal B}(H)\rightarrow{\mathcal B}(H)$ be the linear map
 defined by
$$\Delta (T)=\sum_{k=1}^sA_kTA_k^\dagger+\sum_{i\not= j} E_{ij}TE_{ij}^\dagger -\sum_{l=1}^t
B_lTB_l^\dagger \eqno(5.1)$$ for every $T\in{\mathcal B}(H)$. If
$\sum_{k=1}^s|a_{ki}|^2\geq \sum_{l=1}^t|b_{li}|^2$, $|\sum_{k=1}^s
a_{ki}a_{kj}-\sum_{l=1}^s b_{li}b_{lj}|\leq 1$ whenever $i\not=j$,
then $\Delta $ is NCP positive.}

{\bf Proof.} It is clear that $\Delta$ defined in Eq.(5.1) is not
completely positive since $B_j$ is linearly independent to $\{A_k,
E_{ij}: 1\leq k\leq s; 1\leq i,j\leq n, i\not=j\}$. Assume that
 $\sum_{k=1}^s|a_{ki}|^2\geq \sum_{l=1}^t|b_{li}|^2$, $|\sum_{k=1}^s
a_{ki}a_{kj}-\sum_{l=1}^s b_{li}b_{lj}|\leq 1$ whenever $i\not=j$,
 We will show that $\Delta$ is
 positive.

 Note that
 $$\Delta (E_{mm})=(\sum_{k=1}^s|a_{km}|^2-\sum_{l=1}^t|b_{lm}|^2)E_{kk}+\sum_{i\not=k}E_{ii} \eqno(5.2)$$
 and
 $$\Delta (E_{ij})=(\sum_{k=1}^sa_{ki}\bar{a}_{kj}-\sum_{l=1}^tb_{li}\bar{b}_{lj})E_{ij} \quad{\rm if}\ i\not=j. \eqno(5.3)$$
Let $f_{ii}=\sum_{k=1}^s|a_{ki}|^2-\sum_{l=1}^t|b_{li}|^2$ and
$f_{ij}=\sum_{k=1}^sa_{ki}\bar{a}_{kj}-\sum_{l=1}^tb_{li}\bar{b}_{lj}$
if $i\not=j$. Clearly, $f_{ji}=\bar{f}_{ij}$ for all $i,j$.

Identify $H$ with ${\mathbb C}^n$.
 For any $|\psi\rangle=(\xi_1,\xi_2,\ldots ,\xi_n)^T\in {\mathbb C}^n$,
 consider the rank-one positive matrix
 $|\psi\rangle\langle\psi| =(\xi_i\bar{\xi_j})$. By Eqs.(5.2) and (5.3) we have
 $$\begin{array}{rl} \Delta (|\psi\rangle\langle\psi|)=& \left(\begin{array}{cccc}
 f_{11}|\xi_1|^2
 &f_{12}\xi_1\bar{\xi_2} & \cdots &f_{1n}\xi_1\bar{\xi_n}\\
f_{21}\xi_2\bar{\xi_1} &f_{22}|\xi_2|^2 &\cdots
& f_{2n}\xi_2\bar{\xi_n} \\
\vdots &\vdots &\ddots &\vdots\\ f_{n1}\xi_n\bar{\xi_1}
&f_{n2}\xi_n\bar{\xi_2} & \cdots
&f_{nn}|\xi_n|^2  \end{array}\right) \\
& +
\left(\begin{array}{cccc} \sum_{1\leq j\leq n, j\not=1}|\xi_j|^2 &0&\cdots &0\\
0&\sum_{1\leq j\leq n, j\not=2}|\xi_j|^2 & \cdots &0\\
\vdots &\vdots&\ddots & \vdots \\  0&0& \cdots & \sum_{1\leq j\leq
n, j\not=n}|\xi_j|^2
\end{array}\right)\\
\geq & \left(\begin{array}{cccc}
 \sum_{1\leq j\leq n, j\not=1}|\xi_j|^2
 &f_{12}\xi_1\bar{\xi_2} & \cdots &f_{1n}\xi_1\bar{\xi_n}\\
f_{21}\xi_2\bar{\xi_1} &\sum_{1\leq j\leq n,
j\not=2}|\xi_j|^2&\cdots
& f_{2n}\xi_2\bar{\xi_n} \\
\vdots &\vdots &\ddots &\vdots\\ f_{n1}\xi_n\bar{\xi_1} &
f_{n2}\xi_n\bar{\xi_2}& \cdots &\sum_{1\leq j\leq n,
j\not=n}|\xi_j|^2
\end{array}\right)\\ =&C_{\psi}
\end{array}
$$
So it suffices to show that $C_{\psi}\geq 0$.

To do this, denote $c_i=|\xi_i|$. Then, by the assumption of
$|f_{ij}|\leq 1$ for $i\not=j$, we have
$f_{ij}\xi_i\bar{\xi_j}=c_ic_jv_{ij}$ with $|v_{ij}|\leq 1$, and
$$C_{\psi}=\left(\begin{array}{cccc}
 \sum_{1\leq j\leq n, j\not=1}c_j^2
 &c_1c_2v_{12} & \cdots &c_1c_nv_{1n}\\
c_1c_2\bar{v_{12}} &\sum_{1\leq j\leq n, j\not=2}c_j^2&\cdots
& c_2c_nv_{2n} \\
\vdots &\vdots &\ddots &\vdots\\ c_1c_n\bar{v_{1n}}
&c_2c_n\bar{v_{2n}}& \cdots &\sum_{1\leq j\leq n, j\not=n}c_j^2
\end{array}\right). $$
For any $|\phi\rangle=(\eta_1, \eta_2,\ldots ,\eta_n)^T\in{\mathbb
C}^n$, writing $d_i=|\eta_i|$, we have
$$\begin{array}{rl} \langle \phi |C_\psi\phi\rangle =& \sum_{i=1}^n(\sum_{1\leq
j\leq n, j\not= i}c_j^2)|\eta_i|^2+2{\rm Re}(\sum_{i<j}
c_ic_jv_{ij}\eta_j\bar{\eta_i})\\
\geq & \sum_{i=1}^n(\sum_{1\leq j\leq n, j\not=
i}c_j^2)d_i^2-2\sum_{i<j} c_ic_jd_id_j \\
=&\sum_{i<j}(c_id_j-c_jd_i)^2\geq 0.
\end{array}$$
Therefore, $C_\psi\geq 0$. We have proved that $\Delta
(|\psi\rangle\langle\psi|)\geq 0$ holds for all rank-one positive
matrices $|\psi\rangle\langle\psi|$. It follows that $\Delta$ is a
positive linear map, as desired.\hfill$\square$

The next result gives a NCP positive maps on $3\times 3$ matrices.

{\bf Proposition 5.2.} Let $\Gamma:M_3({\mathbb C})\rightarrow
M_3({\mathbb C})$ be defined by
$$\begin{array}{rl} \Gamma
(A)=&\sum_{i=1}^3E_{ii}AE_{ii}+E_{12}AE_{21}+E_{23}AE_{32}+E_{31}AE_{13}\\
&+\sum_{i\not=j}G_{ij}AG_{ij}^\dag-\sum_{i\not=j}F_{ij}AF_{ij}^\dag
\end{array}\eqno(5.4)$$ for all $A$, where
$F_{ij}=\frac{1}{2}(E_{ii}+E_{jj})$ and
$G_{ij}=\frac{1}{2}(E_{ii}-E_{jj})$, $i,j=1,2,3$ and $i\not=j$. Then
$\Gamma$ is positive and indecomposable.

It is clear that $\Gamma$ is not completely positive by the results
in Section 2. We will show that $\Gamma$ is positive. There is a
easy way to check it. Obviously, $\Gamma$ maps $A=(a_{ij})$ to the
matrix
$$\Gamma(A)=\left( \begin{array}{ccc} a_{11}+a_{22}
&-a_{12}&-a_{13}\\
-a_{21} &a_{22}+a_{33} &-a_{23} \\
-a_{31} &-a_{32} & a_{33}+a_{11}\end{array}\right).$$ So, we need
only check that, if
$$D=\left( \begin{array}{ccc} a
&c&f\\
\bar{c} &b &e \\
\bar{f} &\bar{e} & d\end{array}\right)$$ is positive, then
$$\tilde{D}=\left( \begin{array}{ccc} a+b
&-c&-f\\
-\bar{c} &b+d &-e \\
-\bar{f} &-\bar{e} & d+a\end{array}\right)$$ is positive, and this
  suffices to show that $\det(\tilde{D})\geq 0$. As $D\geq 0$, we
have
$$abd+ce\bar{f}+\bar{c}\bar{e}f-b|f|^2-a|e|^2-d|c|^2\geq 0.$$ Also,
there are numbers $t,s,r\in{\mathbb C}$ with $|t|\leq 1,|s|\leq 1 $
and $|r|\leq 1$ such that $c=\sqrt{ab}t$, $e=\sqrt{bd}s$ and
$f=\sqrt{ad}r$. Thus
$$\det(\tilde{D})\geq a^2d+ab^2+bd^2 +abd\geq 0.$$

The  map $\Gamma$ is also an example that is indecomposable. Recall
that a positive elementary operator decomposable if it has the form
$\Delta_1+\Delta_2^t$ for some completely positive elementary
operators $\Delta_1$ and $\Delta_2$.  Now, for any positive numbers
$a,b$ with $ab\geq 1$, let $\rho\in\mathcal{S}({\mathbb C}^3\otimes
{\mathbb C}^3)$ be
$$\rho=\frac{1}{3(1+a+b)}\left(\begin{array}{ccccccccc}
1&0&0&0&1&0&0&0&1\\
0&a&0&1&0&0&0&0&0\\
0&0&b&0&0&0&1&0&0\\
0&1&0&b&0&0&0&0&0\\
1&0&0&0&1&0&0&0&1\\
0&0&0&0&0&a&0&1&0\\
0&0&1&0&0&0&a&0&0\\
0&0&0&0&0&1&0&b&0\\
1&0&0&0&1&0&0&0&1\end{array}\right)=\frac{1}{3(1+a+b)}\rho_0,$$
where $a\neq1$. $\rho$ is a PPT state since $\rho$ is symmetric
under partial transpose. However,
$$\begin{array}{c}(I_3\otimes\Gamma)(\rho_0)=\left(\begin{array}{ccccccccc}
1+a&0&0&0&-1&0&0&0&-1\\
0&a+b&0&-1&0&0&0&0&0\\
0&0&b+1&0&0&0&-1&0&0\\
0&-1&0&b+1&0&0&0&0&0\\
-1&0&0&0&1+a&0&0&0&-1\\
0&0&0&0&0&a+b&0&-1&0\\
0&0&-1&0&0&0&a+b&0&0\\
0&0&0&0&0&-1&0&b+1&0\\
-1&0&0&0&-1&0&0&0&1+a\end{array}\right)\end{array}$$ is not
positive. Indeed, the vector $(1,0,0,0,1,0,0,0,1)^t$ is an
eigenvector of $(I_3\otimes \Gamma)(\rho_0)$ with eigenvalue $a-1$,
which is negative for $a<1$. This ensures the inseparability of the
PPT state $\rho$. Consequently, $\Gamma$ is indecomposable.

In the case $a>1$, inseparability of $\rho$ can't be detected by the
map $\Gamma$. Let $\Gamma'$ be a linear map defined by
$$\begin{array}{rl} \Gamma'
(A)=&\sum_{i=1}^3E_{ii}AE_{ii}+E_{21}AE_{12}+E_{32}AE_{23}+E_{13}AE_{31}\\
&+\sum_{i\not=j}G_{ij}AG_{ij}^\dag-\sum_{i\not=j}F_{ij}AF_{ij}^\dag
\end{array}\eqno(5.5)$$ for all $A$, where
$F_{ij}=\frac{1}{{2}}(E_{ii}+E_{jj})$ and
$G_{ij}=\frac{1}{{2}}(E_{ii}-E_{jj})$, $i,j=1,2,3$ and $i\not=j$.
$\Gamma'$ is a positive map as well and it maps
$$\left(\begin{array}{ccc}a_{11}&a_{12}&a_{13}\\
a_{21}&a_{22}&a_{23}\\
a_{31}&a_{32}&a_{33}\end{array}\right)
\mapsto\left(\begin{array}{ccc}
a_{11}+a_{33}&-a_{12}&-a_{13}\\
-a_{21}&a_{22}+a_{11}&-a_{23}\\
-a_{31}&-a_{32}&a_{33}+a_{22}\end{array}\right).$$ By a simple
calculation, we get that $(I_3\otimes \Gamma')(\rho_0)$ has a
negative eigenvalue $b-1$ whenever $b<1$. Namely, $\rho$ is detected
by the NCP positive map $\Gamma'$.

Finally, we illustrate that the positive map $\Gamma$ (as well as
$\Gamma^\prime$) can detect some entangled states that cannot be
detected by the realignment criterion. To see this let

$$\rho_1=\frac{1}{195}\left(\begin{array}{ccccccccc}0.99&0&0&0&0.99&0&0&0&0.99\\
0&63&0&0&0&0&0&0&0\\
0&0&1.01&1.01&0&0&0&1.01&0\\
0&0&1.01&1.01&0&0&0&1.01&0\\
0.99&0&0&0&0.99&0&0&0&0.99\\
0&0&0&0&0&63&0&0&0\\
0&0&0&0&0&0&63&0&0\\
0&0&1.01&1.01&0&0&0&1.01&0\\
0.99&0&0&0&0.99&0&0&0&0.99
\end{array}\right). $$
Then $\rho_1$ is   a PPT state. The realignment matrix of $\rho_1$
is
$$\rho_1^R=\frac{1}{195}\left(\begin{array}{ccccccccc}0.99&0&0&0&63&0&0&0&1.01\\
0&0.99&0&0&0&0&1.01&0&0\\
0&0&0.99&0&0&0&0&1.01&0\\
0&0&1.01&0.99&0&0&0&0&0\\
1.01&0&0&0&0.99&0&0&0&63\\
0&1.01&0&0&0&0.99&0&0&0\\
0&0&0&0&0&1.01&0.99&0&0\\
0&0&0&1.01&0&0&0&0.99&0\\
63&0&0&0&1.01&0&0&0&0.99
\end{array}\right). $$
By computation, we have that the trace norm $\|\rho_1^R\|_1\doteq
0.9705<1$. Thus  the realignment criterion does not apply to
$\rho_1$.  Note that,
$$\begin{array}{rl}&(I_3\otimes\Gamma)(\rho_1)\\
=&\frac{1}{195}\left(\begin{array}{ccccccccc}63.99&0&0&0&-0.99&0&0&0&-0.99\\
0&64.01&0&0&0&0&0&0&0\\
0&0&-1.01&2&0&0&0&-1.01&0\\
0&0&-1.01&2&0&0&0&-1.01&0\\
-0.99&0&0&0&63.99&0&0&0&-0.99\\
0&0&0&0&0&64.01&0&0&0\\
0&0&0&0&0&0&64.01&0&0\\
0&0&-1.01&-1.01&0&0&0&2&0\\
-0.99&0&0&0&-0.99&0&0&0&63.99
\end{array}\right).\end{array}$$
The eigenvalues of the last matrix are $$\{-\frac{2}{19500},
\frac{301}{19500},\frac{301}{19500},
\frac{6401}{19500},\frac{6401}{19500},\frac{6401}{19500},
\frac{6201}{19500}, \frac{6498}{19500},\frac{6498}{19500}\}.$$ Thus
$(I_3\otimes\Gamma)(\rho_1)$ has a negative eigenvalue and hence is
not positive. This reveals that $\rho_1$ is entangled, however
cannot be detected by the realignment criterion.

\section{Conclusion}

Let $H$ and $K$ be complex Hilbert spaces of any dimension. The well
known positive map criterion of separability for finite dimensional
quantum systems  has been generalized to infinite dimensional
quantum systems by St${\o}$mer recently which asserts that a state
$\rho$ on $H\otimes K$ is separably if and only if $(\Phi\otimes
I)\rho$ is positive for any normal positive linear maps
$\Phi:{\mathcal B}(H)\rightarrow{\mathcal B}(K)$. However, this
criterion is not practically applied because of the complicacy of
normal positive linear maps. In this paper we give a concrete
characterization of positive completely bounded normal linear maps
by showing that a completely bounded normal linear map $\Phi$ is
positive if and only if it has the form of $ \Phi(X)=\sum
_{i=1}^{\infty }A_{i}XA_{i}^{\dagger}-\sum _{j=1}^{\infty
}C_{j}XC_{j}^{\dagger}$ for all $X$, where $\{C_j\}$ is a
generalized contractive locally linear combination of $\{A_i\}$;
furthermore, $\Phi$ is completely positive if and only if  $\{C_j\}$
is   a generalized contractive linear combination of $\{A_i\}$, and
in turn, if and only if $\Phi$ has the form of $ \Phi(X)=\sum
_{i=1}^{\infty }B_{i}XB_{i}^{\dagger}$. This particularly gives a
characterization of NCP positive elementary operators. Recall that a
linear map $\Phi$ is called an elementary operator if it has the
form of $\Phi(\cdot)=\sum_{i=1}^n A_i(\cdot)B_i$ with $n<\infty$. If
both $H$ and $K$ are finite dimensional, all linear maps
$\Phi:{\mathcal B}(H)\rightarrow{\mathcal B}(K)$ are elementary
operators. Thus an elementary operator $\Phi$ is NCP positive if and
only if there exist $C_{1},\cdots ,C_{k},D_{1},\cdots
,D_{l}\in{\mathcal B}(H,K)$ such that  $\Phi
(X)=\sum_{i=1}^kC_iXC_i^\dagger-\sum_{j=1}^lD_jXD_j^\dagger$ for all
$X\in{\mathcal B}(H)$, and $\{D_j\}_{j=1}^l$ is a contractive
locally linear combination  but not a contractive linear combination
of $\{C_i\}_{i=1}^k$. Therefore, for elementary operators, the
question when positivity ensures complete positivity may be reduced
to the question when contractive locally linear combination implies
linear combination. This connection allows us to look more deeply
into the relationship and the difference between positivity and
complete positivity, and obtain some simple criteria to check
whether a positive elementary operator is completely positive or
not. This is important especially when we construct positive maps
and apply them to recognize entanglement.

Above characterization of positive maps allows us to get a concrete
representation of quantum channels for infinite dimensional systems
that is similar to finite dimensional case. Every Quantum channel
$\mathcal E$ has the form of $ {\mathcal E}(\rho)=\sum
_{i=1}^{\infty }M_{i}\rho M_{i}^{\dagger}$, where $\sum_{i=1}^\infty
M_i^\dagger M_i=I$.

Much more importantly, our characterization of positive maps leads
to  a necessary and sufficient condition of separability which we
call the elementary operator criterion: $\rho$ is separable if and
only if $(\Phi\otimes I)\rho \geq 0$ holds for every positive
 elementary operator $\Phi :{\mathcal B}(H)\rightarrow
{\mathcal B}(K)$, and in turn, if and only if $(\Phi\otimes I)\rho
\geq 0$ holds for every finite-rank positive
 elementary operator $\Phi :{\mathcal B}(H)\rightarrow
{\mathcal B}(K)$. Equivalently, $\rho$ is entangled if and only if
there exists an elementary operator of the form $\Phi(\cdot )= \sum
_{i=1}^{k}C_{i}(\cdot )C_{i}^{\dagger}-\sum _{j=1}^{l}D_{j}(\cdot
)D_{j}^{\dagger}:{\mathcal B}(H)\rightarrow {\mathcal B}(K)$, where
 all $C_i$s and $D_j$s are of finite rank and $\{ D_1,\ldots , D_l\}$ is a
contractive locally linear combination of $\{C_1, \ldots ,C_k\}$,
such that the operator $(\Phi\otimes I)\rho $ is not positive.
Obviously, this criterion is more practical than the positive map
criterion and the St$\o$mer's theorem. Some examples of finite rank
NCP positive elementary operators are given which are used to detect
some entangled states that can not be recognized by the PPT
criterion and the realignment criterion.

{\bf Acknowledgement.} The authors wish to give their thanks to the
referees for  helpful comments and suggestions to improve the
original manuscript.


\begin{thebibliography}{99}




\bibitem{BZ} Bengtsson I, Zyczkowski K,
Cambridge University Press, Cambridge, 2006.

\bibitem{NC} Nielsen M  A,  Chuang I L, Quantum Computation and
Quantum Information, Cambridge University Press, Cambridge, 2000.


\bibitem{W} Werner R  F,
Phys. Rev. A 40 (1989) 4277.


\bibitem{BB} Bennett C H, Brassard G, Cr$\acute{e}$peau C, Jozsa R, Peres A,
Wootters W K, Phys. Rev. Lett. 70 (1993) 1895.

\bibitem{BP} Bouwmeester D, Pan J W, Mattle K, Eibl M, Weinfurter H,
Zeilinger A, Nature 390 (1997) 575.

\bibitem{DE} Deutsch D, Ekert A, Jozsa R,  Macchiavello C, Popescu S,
Sanpera A, Phys. Rev. Lett. 77 (1996) 2818.

\bibitem{DE1} Deutsch D, Ekert A, Jozsa R, Macchiavello C, Popescu S,
A. Sanpera, Phys. Rev. Lett. 80 (1998) 2022.

\bibitem{S} Shor P W, Phys. Rev. A 52 (1995) 2493.

\bibitem{Hor} Horodecki M, Horodecki P, Horodecki R,
Phys. Lett. A, 223 (1996), 1-8.


\bibitem{Pe} Peres A,
Phys. Lett. A 202 (1996) 16.


\bibitem{HHH} Horodecki M, Horodecki P, Horodecki R,
Phys. Lett. A 80 (1998) 5239.


\bibitem{CW} Chen K, Wu L A, Quant. Inf. Comput 3 (2003) 193.

\bibitem{Hor2} Horodecki M, Horodecki P,
Phys. Rev. A, 59 (1999), 4206.

\bibitem{CAG} Cerf N J, Adami C, Gingrich R M,
Phys. Rev. A 60 (1999): 893.

\bibitem{HHH1} Horodecki R, Horodecki P, Horodecki M,
Rev. Mod. Phys. 81 (2009) 865.

\bibitem{HQ}  Hou J C,  Qi X F,
Phys. Rev. A 81 (2010) 062351.

\bibitem{SS} Salgado D,  S\'{a}nchez-G\'{o}mez J L,
Open Systems and Information Dynamics, 12(1) (2005), 55-64.

\bibitem{SSZ} Skowronek {\L}, St{\o}rmer E,  \.{Z}yczkowski K,
J. Math. Phys.
50, 062106 (2009); doi:10.1063/1.3155378

\bibitem{GKM}  Grabowski J,  Ku\'{s} M,  Marmo G,
Open. Sys. and Information Dyn.
14 (2007), 355-370.

\bibitem{K} Kye S-H,
Trends in Mathematics, Information Center for Mathematical Sciences,
6(2) (2003), 83-91.


\bibitem{Kad}  Kadison R V,  Ringrose J R, Fundamentals of the
Theorey of Operator Algebras II, Graduate Studies in Math., 16,
American Math. Society, New York: Academic Press, 1983.

\bibitem{Sto}  St${\o}$rmer E,
J. Func. Anal., 254 (2008), 2304-2313.

\bibitem{Dix}  Dixmier J, Von Neumann Algebras, North-Holland
Publishing Com., Amsterdan, New York, Oxford, 1981.

\bibitem{CK}   Chru$\acute{s}$ci$\acute{n}$ski D,  Kossakowski A,
Open Systems and Inf. Dynamics 14 (2007) 275;

\bibitem{CK2}  Chru$\acute{s}$ci$\acute{n}$ski D, Kossakowski A, J. Phys. A:
Math. Theor. 41 (2008) 145301.

\bibitem{AS} Augusiak R, Stasi¡änska J, Phys. Rev. A 77 (2008)
010303.


\bibitem{C1}  Choi M D,
Lin. Alg. Appl., 10(1975), 285-290.

\bibitem{C2}  Choi M D,
J. Operator Theory, 4(1980), 271-285.

\bibitem{C3}  Choi M D,
Proc. Sympos. Pure
Math., 38(1982), 583-590.

\bibitem{De} Depillis J,
Pacific J. Math., 23 (1967),
129-137.

\bibitem{H4}  Hou J C,
J. Operator Theory,
39 (1998), 43-58.

\bibitem{P} Paulsen V, Completely Bounded Maps and Operator Algebras, Cambridge Studies in
Advanced Mathematics 78, Cambridge University Press, Cambridge,
2002.

\bibitem{St}  Stinespring W F,
Proc. Amer. Math. Soc., 6 (1955), 211-216.

\bibitem{H1}  Hou J C,
Sci. in China (ser.A), 32(1989), 929-940.

\bibitem{M2}  Mathieu M,
Math. Ann., 284(1989), 223-244.

\bibitem{ER}  Effros E G,  Ruan Z-J, Operator Spaces, Clarendon
Press, Oxford, 2000.

\bibitem{H5} Hou J C,
Sci. in China (ser.A), 36(9) (1993), 1025-1035.


\bibitem{SV2}  Sperling J,  Vogel W,
Phys. Rev. A 79 (2009) 052313.


\end{thebibliography}
\end{document}